\documentclass[aps,pra,twocolumn,showpacs]{revtex4}

\usepackage{color, graphicx}
\usepackage{amsmath, amssymb}

\begin{document}

\title{Quantum state engineering by click counting}

\author{J. Sperling}\email{jan.sperling@uni-rostock.de}\affiliation{Arbeitsgruppe Theoretische Quantenoptik, Institut f\"ur Physik, Universit\"at Rostock, D-18051 Rostock, Germany}
\author{W. Vogel}\affiliation{Arbeitsgruppe Theoretische Quantenoptik, Institut f\"ur Physik, Universit\"at Rostock, D-18051 Rostock, Germany}
\author{G. S. Agarwal}\affiliation{Department of Physics, Oklahoma State University, Stillwater, Oklahoma 74078, USA}

\date{\today}

\begin{abstract}
	We derive an analytical description for quantum state preparation using systems of on-off detectors.
	Our method will apply the true click statistics of such detector systems.
	In particular, we consider heralded quantum state preparation using correlated light fields, photon addition, and photon subtraction processes.
	Using a post-selection procedure to a particular number of clicks of the detector system, the output states reveal a variety of quantum features.
	The rigorous description allows the identification and characterization of fundamentally unavoidable attenuations within given processes.
	We also generalize a known scenario of noiseless amplification with click detectors for the purpose of the preparation of various types of nonclassical states of light.
	Our exact results are useful for a choice of experimental parameters to realize a target state.
\end{abstract}

\maketitle


\section{Introduction}

	The measurement and generation of photons is one of most challenging tasks in nowadays quantum optics.
	Since the photon can be considered as the carrier of information, it plays a fundamental role in quantum information and quantum communications~\cite{KMNRDM07}.
	The generation of single photons is typically described by conditional measurements of quantum correlated light fields with single photon detectors~\cite{HM86,JHBL10,WLRBGCZCP10,SCSWS11,AMFL13,BDSJBDSW12,CXRVHSRKSCE13,FFWSACSLM13}.
	The sophisticated task is to find a proper device that can detect -- at least in a good approximation -- single photons.
	Among other approaches, e.g.,~\cite{BC10,SSTT13,MVSHLGVBSMN13,NPGGSBKALSLS14}, an avalanche photo diode in Geiger mode is an experimentally accessible device, being close to a single photon counter~\cite{ABA10,DYSTS11}.
	However, this detector not only produces a click signal in the case of a single incident photon, but also when multiple photons have been absorbed.

	A possible way to partially overcome this ambiguity is given by a joint measurement of a signal with multiple click detectors.
	One implementation is given by so-called multiplexing detection schemes, see, e.g.,~\cite{ASSBW03,FJPF03,ZABGGBRP05,FLCEPW09}, and another one employs detector arrays, see, e.g.,~\cite{WDSBY04,JDC07,DBIMHCE13}.
	In both scenarios, the incident light is split equally into $N$ modes, and each mode can be measured with one of those on-off diodes.
	The probability $c_k$ for a total number of $k$ clicks is described through the click counting distribution~\cite{SVAPRA12},
	\begin{align}\label{Eq:ClickStatistics}
		c_k=\left\langle{:}\binom{N}{k}\left({\rm e}^{-\frac{\eta\hat n}{N}}\right)^{N-k}
		\left(\hat 1-{\rm e}^{-\frac{\eta\hat n}{N}}\right)^{k}{:}\right\rangle,
	\end{align}
	with $k=0,\ldots,N$ denoting the number of clicks, $N$ being the number of on-off detectors, $\eta$ the quantum efficiency, and ${:}\,\,\,{:}$ denotes the normal ordering prescription.
	Based on the variance of these statistics, it is possible to identify in theory and experiment nonclassical, i.e. sub-binomial, light~\cite{SVAPRL12,BDJDBW13}.

	A source of correlated photon pairs together with the considered class of detectors can be used, for example, to identify spatial correlations~\cite{BDFL08,MMDL12,SVAPRA13} or to perform a detector calibration~\cite{PHMH12,PHMH13}.
	Moreover, these states 	allow one to predict the presence of a photon in one mode, if a photon is detected in the other one.
	For the first time, this so-called heralded generation of single photons has been experimentally realized in Ref.~\cite{HM86}.
	Recent developments led to a further enhancement of this kind of single-photon source~\cite{JHBL10,WLRBGCZCP10,SCSWS11,AMFL13,BDSJBDSW12,CXRVHSRKSCE13,FFWSACSLM13}.
	Other protocols to manipulate quantum states in theory and experiment in the single photon regime are known as photon addition and photon subtraction~\cite{AT92,ZVB04,BA07,KVPZB08,DM09,KVBZ11,LLNJ12,FMCZB13,RKVGZB13}.
	These engineering protocols can be also used to add or subtract several photons~\cite{MF10,F09} or to probe experimentally fundamental commutation relations~\cite{PZKB07}.
	Another prominent example for controlled state manipulation is noiseless amplification~\cite{C82,RL09,FBBFTG10,ZFB11,BPMT13}.
	All these engineering processes are able to enhance quantum properties, such as entanglement, for applications in quantum information science, see, e.g.,~\cite{MD02,BCZW10,HHFRRJ10,XRLWP10,LN12,NGSC12,OBSZGT12,AN13,GLS13,MBHSAGLS13,SHRSHB13}.

	In the present contribution, we study the quantum state engineering for conditional measurements with on-off detector systems.
	We will consider three schemes.
	First, the conditional measurement in one mode of a quantum correlated bipartite states.
	Second, a multi-photon subtraction protocol, and, third, a multiple photon addition process.
	In all scenarios, we consider imperfect detectors for a realistic description of the underlying physical situation.
	As an example, we study a noiseless amplification scenario and apply it to the conversion of coherent light into different types of nonclassical states.

	The paper is structured as follows.
	In Sec.~\ref{Sec:Heralding}, we consider the heralded quantum state manipulation by measurements with on-off detector systems.
	In Secs.~\ref{Sec:Subtraction} and~\ref{Sec:Addition}, we describe the multi-photon subtraction and addition processes, respectively, applying the same detection process.
	A combination of addition and subtraction yields the noiseless amplification procedure, which is applied in Sec.~\ref{Sec:NA} to the engineering of nonclassical states.
	We conclude in Sec.~\ref{Sec:Conclusions}.


\section{Heralded state preparation with click detector systems}\label{Sec:Heralding}

\begin{figure}[ht]
	\includegraphics*[width=4.2cm]{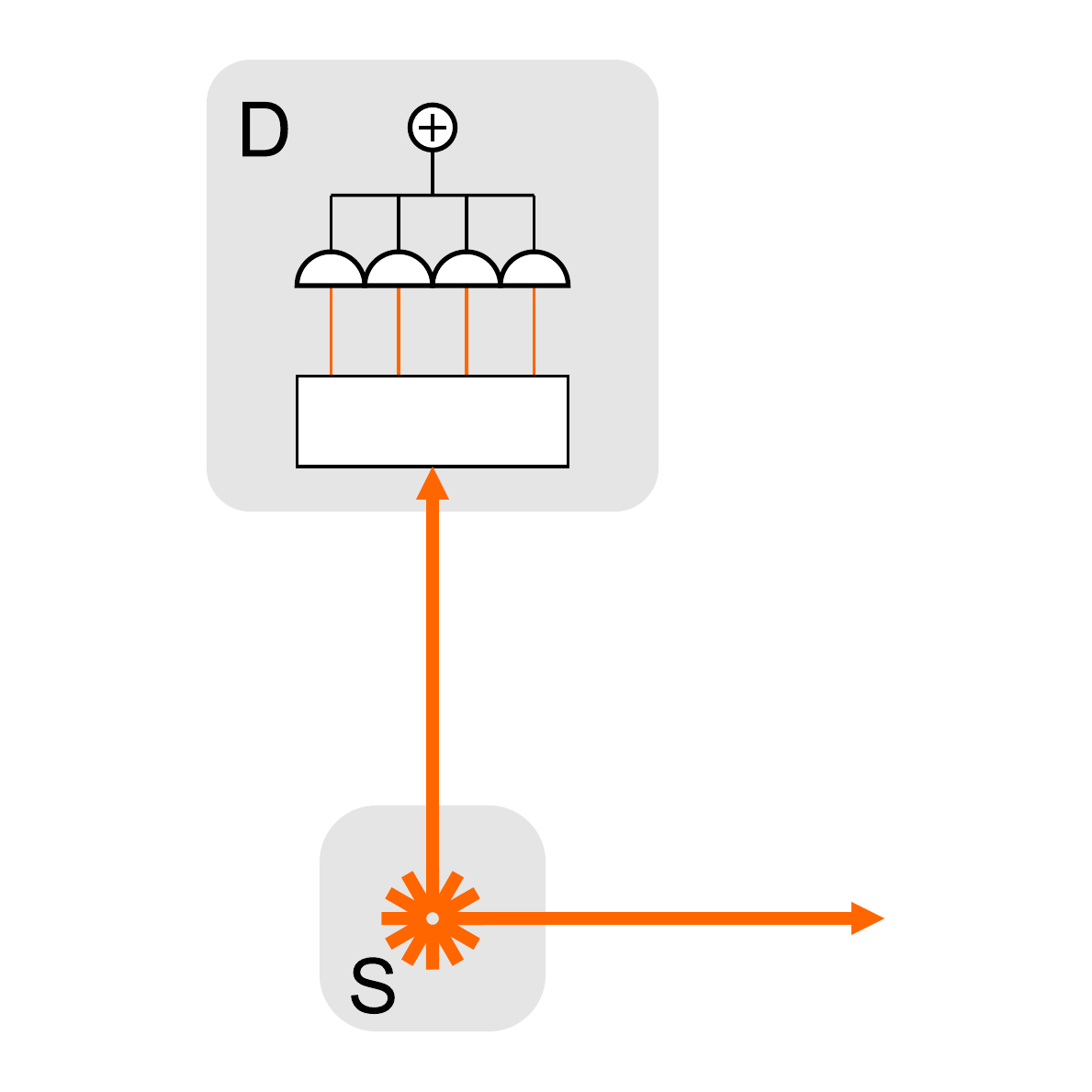}
	\caption{
		(Color online)
		A source S generates quantum correlated light.
		One beam is measured with a click detector system D.
		The other beam will be further processed only in case of a $k$ click event of the detector.
	}\label{Fig:HeraldingScheme}
	\end{figure}

	Let us start with the heralding scheme, see Fig.~\ref{Fig:HeraldingScheme}.
	A quantum correlated light field may be generated by a parametric process in order to certify a photon-photon correlation between two modes: $A$ and $B$.
	In this scenario, one part of the radiation field (mode $B$) is measured with a click detector system.
	In such a detector the incident light beam is split into $N$ output states with equal intensities.
	Each of the resulting beams is measured with a single avalanche photo diode.
	The joint number of clicks $k$ yields the click counting statistics $c_k$ in Eq.~\eqref{Eq:ClickStatistics}, for $k=0,\ldots,N$.
	The remaining part of the field (mode $A$) is further processed, solely if $k$ clicks have been measured in mode $B$.

	Before we start, let us briefly review some properties of a click counter as a photon number resolving device.
	The click counting statistics in Eq.~\eqref{Eq:ClickStatistics} is the quantum analog to the binomial statistics.
	Whereas the true photon statistics of a quantum state $\hat\rho$ is given by
	\begin{align}
		p_k=\langle k|\hat\rho|k\rangle=\left\langle{:}\frac{\hat n^k}{k!}{\rm e}^{-\hat n}{:}\right\rangle,
	\end{align}
	for $k=0,1,2,\ldots$ -- resembling a Poissonian form~\cite{BookVogel,BookAgarwal}.
	In previous works, we have shown that the binomial form requires a reformulation of nonclassicality conditions~\cite{SVAPRL12,SVAPRA13}.
	Moments of the click counting statistics can be used to identify quantum correlations -- even between multiple detector systems.
	Moreover, an extension to general photon absorption processes led to the description of on-off diodes working in a non-linear interaction regime, e.g., two-photon absorption.
	Similarly, it allows the description of noise models, e.g., dark counts.
	It is worth mentioning that the click counting statistics approaches the photoelectric counting theory in the limit of an infinite number of avalanche diodes, $N\to\infty$.

	Since we aim to describe protocols for quantum state engineering, it is important to describe the click counting statistics in terms of a positive operator valued measure (POVM).
	In general, the POVM element, with $c_k=\langle\hat \Pi_k\rangle$ for $0\leq\eta\leq1$, can be written as
	\begin{align}
		\nonumber \hat \Pi_k=&{:} \binom{N}{k} \left({\rm e}^{-\frac{\eta\hat n}{N}}\right)^{N-k}\left(\hat 1-{\rm e}^{-\frac{\eta\hat n}{N}}\right)^{k}{:}
		\\ =&\binom{N}{k}\sum_{j=1}^k\binom{k}{j}(-1)^{k-j}{:}{\rm e}^{-\eta\left(1-\frac{j}{N}\right)\hat n}{:}\,.
		\label{Eq:ClickPOVM}
	\end{align}
	Hence, all POVM elements can be written as a linear combination of the operator
	\begin{align}\label{Eq:SimpleMeasure}
		\hat M(\lambda)={:}{\rm e}^{-\lambda\hat n}{:}=\sum_{m=0}^\infty (1-\lambda)^m |m\rangle\langle m|,
	\end{align}
	for different values of $\lambda$.
	Using the $\mathcal D$-symbol, cf. Appendix~\ref{App:Dsym}, we get
	\begin{align}\label{Eq:DsymClickPOVM}
		\hat \Pi_k=\sum_{m=0}^\infty \mathcal D^{1-\eta,\eta}_{k,m} |m\rangle\langle m|.
	\end{align}
	In the following, we are going to use this representation for conditional measurements.
	A general, bipartite input state $\hat\rho_{\rm in}$ is given in Fock basis expansion as
	\begin{align}
		\hat\rho_{\rm in}=\sum_{p,q,r,s=0}^\infty\rho_{p,q,r,s}|p\rangle\langle q|\otimes|r\rangle\langle s|.
	\end{align}
	Using the POVM element $\hat\Pi_k$ for $k$ clicks, we find
	\begin{align}
		\nonumber \hat\rho_{k,\rm out}=&{\rm tr}_B\left(\hat\rho_{\rm in}\left[\hat 1\otimes\hat\Pi_k\right]\right)\\
		=&\sum_{p,q=0}^\infty\left[\sum_{m=k}^\infty \mathcal D^{1-\eta,\eta}_{k,m}\rho_{p,q,m,m}\right]|p\rangle\langle q|.
		\label{Eq:HeraldingOUT}
	\end{align}
	This result already represents the output state -- triggered to $k$ clicks -- of the scheme in Fig.~\ref{Fig:HeraldingScheme} for $N$ diodes, including the quantum efficiency $\eta$.

\begin{figure}[ht]
	\includegraphics*[width=8.4cm]{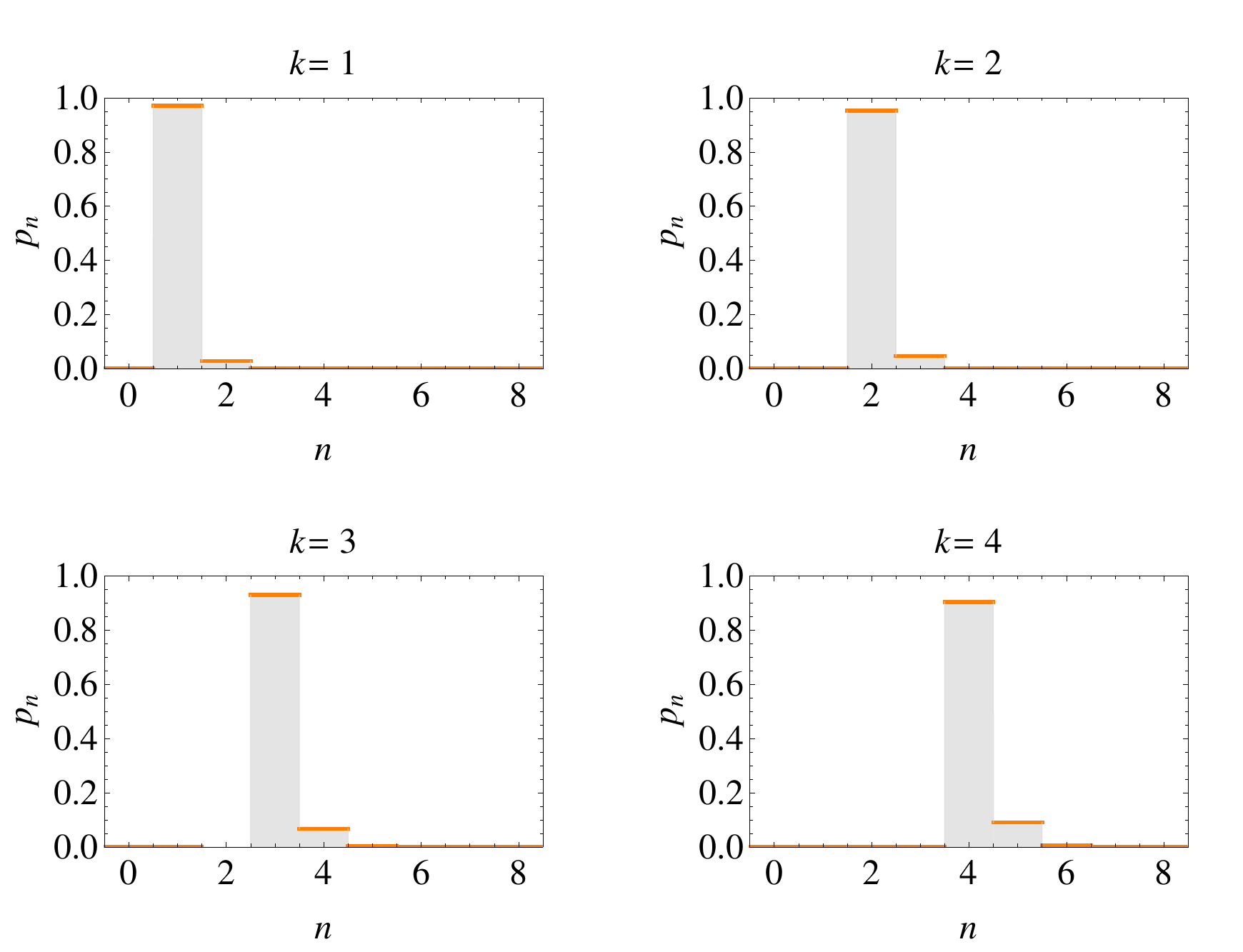}
	\caption{
		(Color online)
		The output photon distributions $p_n$ is shown for the photon-photon correlated input in Eq.~\eqref{Eq:ASVstate} ($\omega=0.25$) triggered to $k$ clicks.
		For a small number of clicks, $k\ll N=64$, and a high quantum efficiency, $\eta=0.95$, the heralding yields a good estimate to a $k$-photon state.
	}\label{Fig:Heralding}
	\end{figure}

	As an example let us consider a two-mode squeezed-vacuum state undergoing a full phase diffusion,
	\begin{align}\label{Eq:ASVstate}
		\hat\rho_{\rm in}=(1-\omega)\sum_{n=0}^\infty \omega^n |n,n\rangle\langle n,n|,
	\end{align}
	with $0<\omega<1$.
	Interestingly, although this state is considered to be classically correlated with respect to a number of notions of quantumness, it has been shown to be two-mode quantum correlated~\cite{ASV13}.
	In Fig.~\ref{Fig:Heralding}, we show the photon statistics $p_n$ of the heralded output states in Eq.~\eqref{Eq:HeraldingOUT}, depending on the number of clicks $k$ of the detector,
	\begin{align}
		p_n=&\langle n|\hat\rho_{k,\rm out}|n\rangle
		=(1-\omega)\omega^n\mathcal D^{1-\eta,\eta}_{k,n}.
	\end{align}
	The aim to generate a $k$-photon state by a post-selection is properly approximated even with a finite quantum efficiency, $\eta<1$.
	Note that the photon distribution is normalized to one for each number of clicks $k$.
	The contribution of higher photon numbers is not negligible for larger values of $k$, however it can be reduced by increasing the number of diodes.
	In the limit, $N\to\infty$ and $\eta\to1$, the output state would be a perfect $k$-photon state, cf. Appendix~\ref{App:ErrE}.

\section{Multi-Photon Subtraction}\label{Sec:Subtraction}

\begin{figure}[ht]
	\includegraphics*[width=4.2cm]{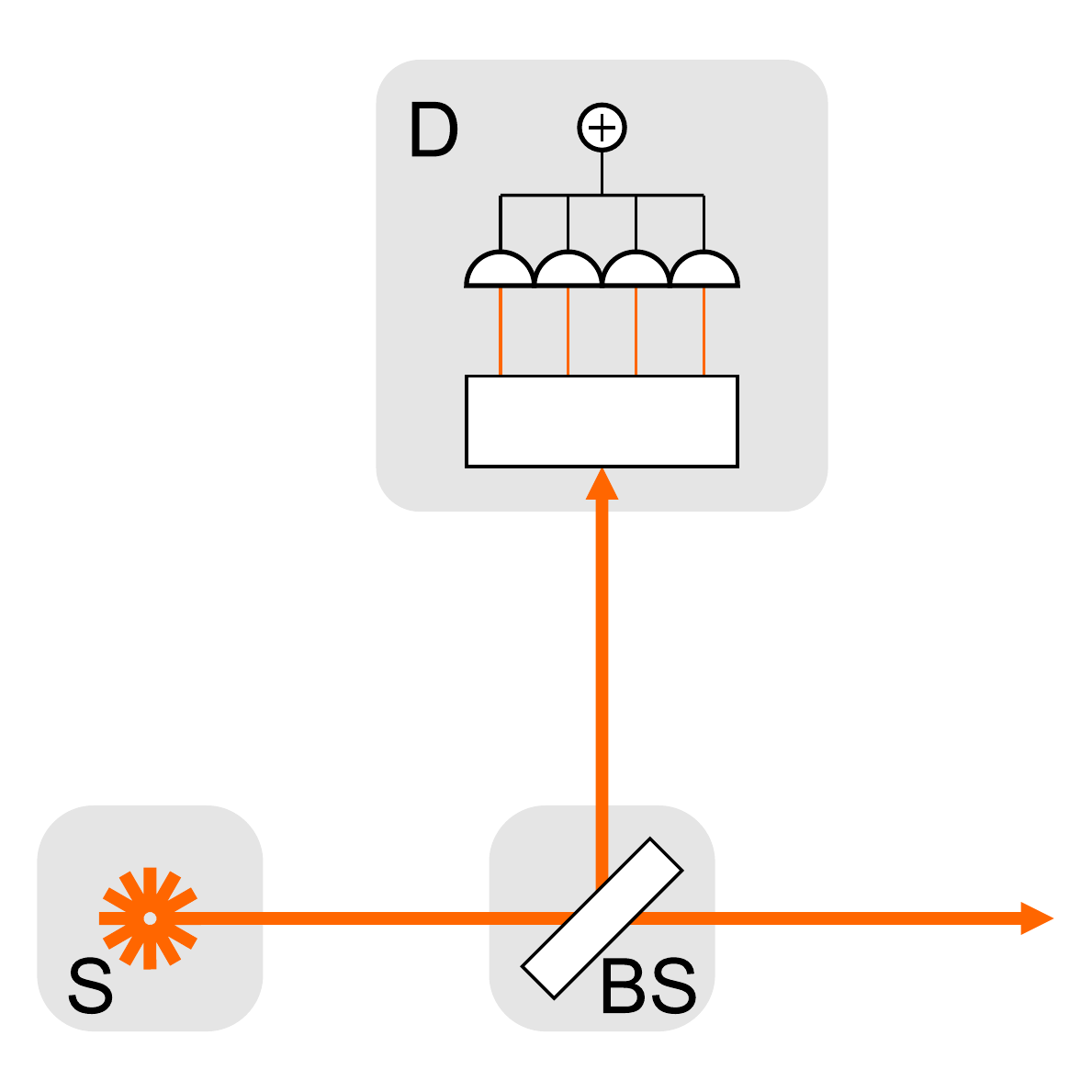}
	\caption{
		(Color online)
		An incident light beam is split on a beam splitter into two outgoing beams.
		One of them is detected with an on-off detector system.
		Only states that correspond to $k$ clicks of the detector will be further processed.
	}\label{Fig:SubtractionScheme}
	\end{figure}

	The idea of a single-photon subtraction is the implementation of the operation:
	\begin{align}
		\hat\rho_{\rm in}\mapsto\hat\rho_{\rm out}=\frac{\hat a\hat\rho_{\rm in}\hat a^\dagger}{{\rm tr}\left[\hat a\hat\rho_{\rm in}\hat a^\dagger\right]}.
	\end{align}
	Such an operation is a classical one, cf., e.g.,~\cite{GSV12}.
	Applying this operation $k$ times would yield a $k$-photon subtraction.
	We aim to formulate this operation within a single detection process with detector systems of multiple on-off diodes.
	For this reason, we study the scenario in Fig.~\ref{Fig:SubtractionScheme}.
	A light beam enters one input of a beam splitter -- vacuum is supposed at the other input port.
	The transmissivity and reflectivity of the beam splitter are $t$ and $r$, respectively, with $t^2+r^2=1$.
	Note that, without loss of generality, $t$ and $r$ can be chosen to be positive real numbers.
	An on-off detector system is applied to measure one of the outputs (mode $B$).
	We post-select those output states in mode $A$, which corresponds to $k$ clicks of the detector.

	Let us start with a coherent input field $|\alpha\rangle$.
	The beam splitter transforms the input field as
	\begin{align}
		\hat a&\mapsto  r\hat a+t \hat b,
	\intertext{which yields an output of}
		|\alpha,0\rangle&\mapsto  |\phi\rangle=|t\alpha,r\alpha\rangle.
	\end{align}
	We perform a partial trace with $\hat M(\lambda)$, cf.~Eq.~\eqref{Eq:SimpleMeasure}, in the second mode.
	Hence, we get a resulting state as
	\begin{align}
		\nonumber &{\rm tr}_B\left(|\phi\rangle\langle \phi|\left[\hat 1\otimes{:}{\rm e}^{-\lambda\hat b^\dagger\hat b}{:}\right]\right)
		={\rm e}^{-\lambda r^2|\alpha|^2}|t\alpha\rangle\langle t\alpha|
		\\=&\sum_{m=0}^\infty \frac{1}{m!}\left(-\lambda\frac{r^2}{t^2}\right)^m \hat a{}^m|t\alpha\rangle\langle t\alpha|\hat a^\dagger{}^m.
		\label{Eq:SubtractionCoherent}
	\end{align}
	Let us note that the scaling $t$ of the coherent amplitude $\alpha$ can be understood as a loss process.

	A general input state $\hat\rho_{\rm in}$ may be given in the Glauber-Sudarshan representation~\cite{G63,S63} as
	\begin{align}
		\hat\rho_{\rm in}=\int\!{\rm d}^2\alpha\,P(\alpha)\,|\alpha\rangle\langle\alpha|,
	\end{align}
	then a loss operation $\Lambda^{\rm (loss)}_{t}$ can be written as
	\begin{align}
		\Lambda^{\rm (loss)}_{t}(\hat\rho_{\rm in})=\int\!{\rm d}^2\alpha\,P_t^{\rm (loss)}(\alpha)\,|\alpha\rangle\langle \alpha|,
	\end{align}
	with the scaled $P$~function:
	\begin{align}
		P_t^{\rm (loss)}(\alpha)=\frac{1}{t^2}P\!\left(\frac{\alpha}{t}\right).
	\end{align}
	Using the decomposition in Eq.~\eqref{Eq:ClickPOVM} together with Eq.~\eqref{Eq:SubtractionCoherent}, we get for the conditional measurement of $k$ clicks, $\hat\Pi_k$, the output state:
	\begin{align}
		\nonumber \hat\rho_{k,\rm out}=&\sum_{m=0}^\infty \frac{1}{m!}\left[\binom{N}{k}\sum_{j=0}^k\binom{k}{j}(-1)^{k-j}\right.
		\\\nonumber &\times\left.\left(-\eta\left[1-\frac{j}{N}\right]\frac{r^2}{t^2}\right)^m\right] \hat a{}^m\Lambda^{\rm (loss)}_{t}(\hat\rho_{\rm in})\hat a^\dagger{}^m
		\\=&\sum_{m=k}^\infty \frac{1}{m!}\mathcal D_{k,m}^{-\eta',\eta'}\hat a{}^m\Lambda^{\rm (loss)}_{t}(\hat\rho_{\rm in})\hat a^\dagger{}^m,
		\label{Eq:SubtractionKrauss}
	\end{align}
	with an effective efficiency of
	\begin{align}
		\eta'=\eta\frac{r^2}{t^2},
	\end{align}
	where we used the property of the $\mathcal D$-symbol that it vanishes, $\mathcal D_{k,m}^{-\eta',\eta'}=0$, for $m<k$, cf. Appendix~\ref{App:Dsym}.
	The process in Eq.~\eqref{Eq:SubtractionKrauss}, $\hat\rho_{\rm in}\mapsto\hat\rho_{k,\rm out}$, describes the $k$ click conditioned measurement protocol in Fig.~\ref{Fig:SubtractionScheme}.
	The output state, $\hat\rho_{k,\rm out}$, is normalized to the probability ${\rm tr}\,\hat\rho_{k,\rm out}$ to have $k$ clicks.

	Let us note that the process in Eq.~\eqref{Eq:SubtractionKrauss} not only subtracts $k$ photons.
	It also includes higher numbers of subtractions.
	Moreover, the output state undergoes a loss process $\Lambda^{\rm (loss)}_t$.
	This feature is a consequence of the process itself and not a result of the imperfect detector, since it is independent of the quantum efficiency $\eta$.

\begin{figure}[ht]
	\includegraphics*[width=8.4cm]{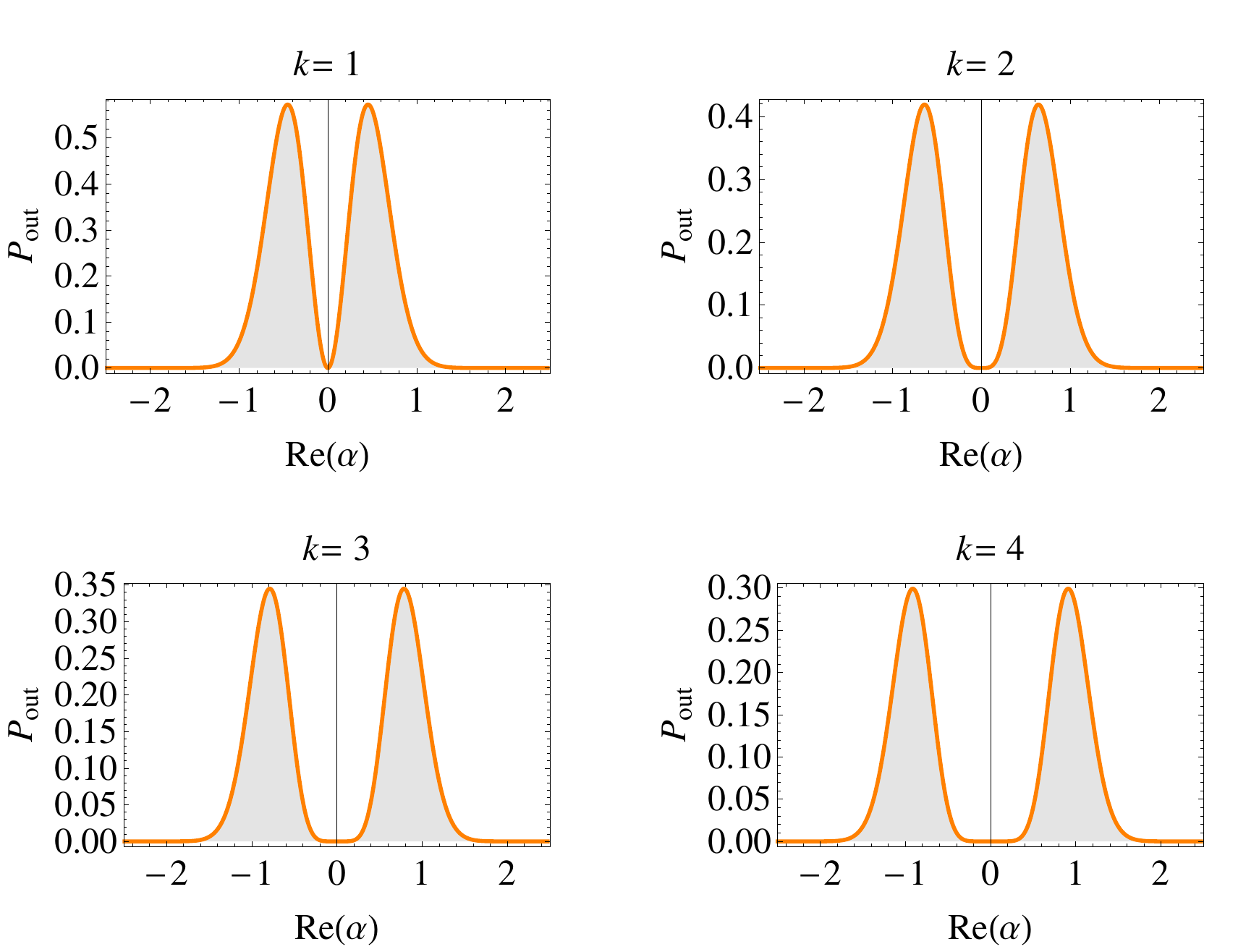}
	\caption{
		(Color online)
		The figures show the output $P$~function of a $k$-click subtraction process for $N=16$ on-off diodes with a quantum efficiency $\eta=0.8$.
		The transmission is $t=0.7$, the mean thermal photon number of the input state is $\bar n=0.5$.
	}\label{Fig:SubtractionExample}
	\end{figure}

	In Fig.~\ref{Fig:SubtractionExample}, we study this quantum process for a thermal input field,
	\begin{align}
		\hat\rho_{\rm in}=&\int\!{\rm d}^2\alpha \frac{{\rm e}^{-|\alpha|^2/\bar n}}{\pi\bar n} |\alpha\rangle\langle\alpha|,
		\label{Eq:ThermalState}
	\end{align}
	with a mean photon number $\bar n$.
	Applying the loss process $\Lambda^{\rm (loss)}_t$ is equivalent to a scaling of the mean photon number: $\bar n_0=t^2\bar n$.
	The output $P$~function triggered to a $k$ click event is
	\begin{align}
		\nonumber P_{\rm out}(\alpha)=&\binom{N}{k}\left({\rm e}^{-\eta'|\alpha|^2/N}\right)^{N-k}
		\\&\times\left(1-{\rm e}^{-\eta'|\alpha|^2/N}\right)^{k}\frac{{\rm e}^{-|\alpha|^2/\bar n_0}}{\pi\bar n_0},
	\end{align}
	cf. Appendix~\ref{App:DThS}, where we additionally give analytical expressions for displaced thermal states.
	The plots of $P_{\rm out}$ in Fig.~\ref{Fig:SubtractionExample} are additionally normalized to one, $\int\!{\rm d}^2\alpha\, P_{\rm out}(\alpha)=1$.
	The number of individual click diodes, $N=16$, corresponds to a $4\times 4$ array detector or to a multiplexing detector with a depth $d=4$, $N=2^d$, with a quantum efficiency of $80\%$.
	Since the considered process is a classical one, the Gaussian input state is deformed into another classical (non-negative) output distribution.

\section{Multi-Photon Addition}\label{Sec:Addition}

	So far, we adapted the photon-subtraction protocol for measurements with on-off detector systems.
	Another frequently studied operation is a single-photon addition:
	\begin{align}
		\hat\rho_{\rm in}\mapsto\hat\rho_{\rm out}=\frac{\hat a^\dagger\hat\rho_{\rm in}\hat a}{{\rm tr}\left[\hat a^\dagger\hat\rho_{\rm in}\hat a\right]}.
	\end{align}
	The photon addition is known to be a nonclassical process and, hence, can be used to generate nonclassical output states from classical inputs, cf., e.g.,~\cite{RKVGZB13}.
	Now, let us study this operation for click counting devices.

\begin{figure}[ht]
	\includegraphics*[width=4.2cm]{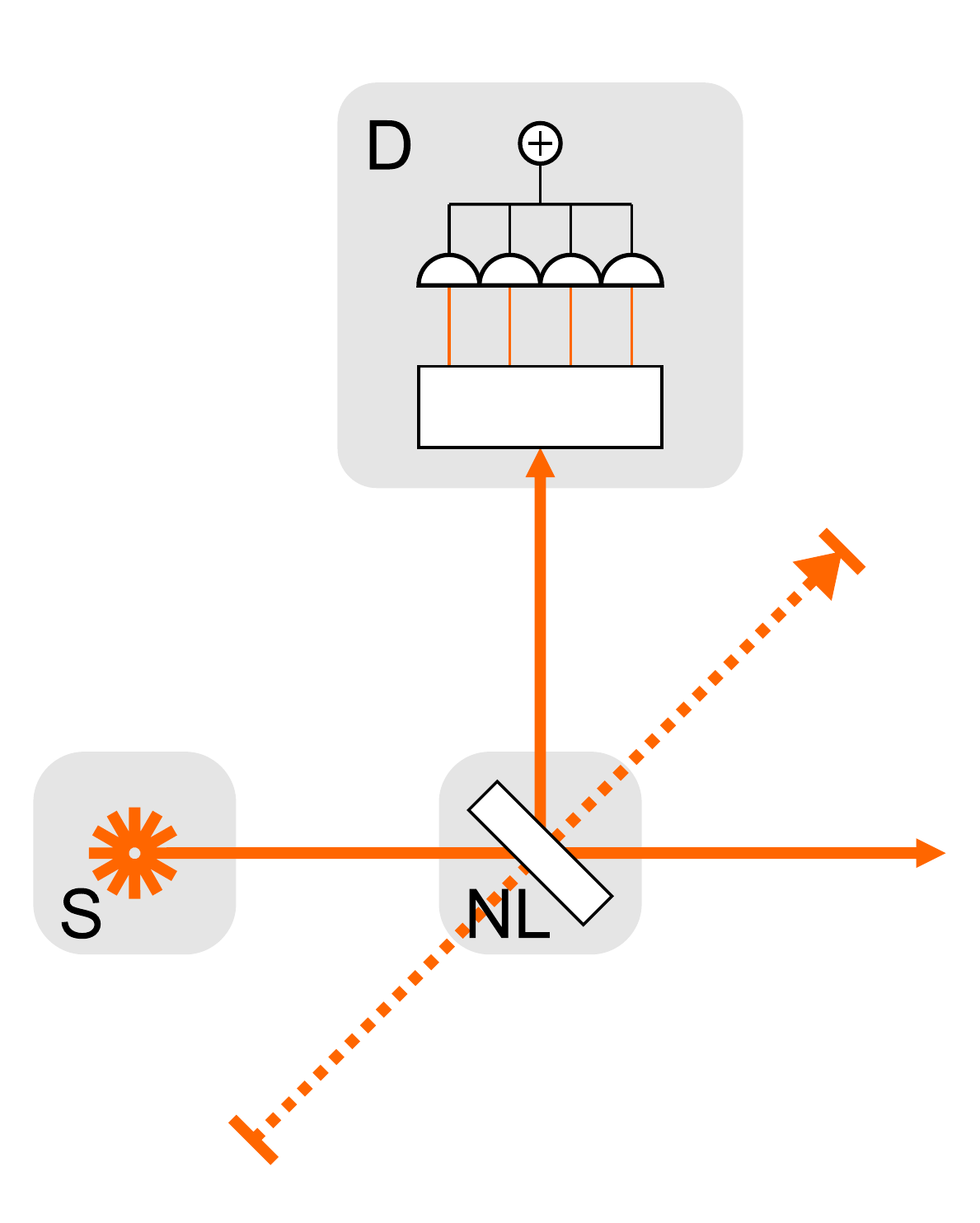}
	\caption{
		(Color online)
		A non-linear crystal (NL) is pumped by a pump beam (dashed line).
		An incident signal field and vacuum are mixed in this non-linear medium.
		One of the generated output beams is measured with a click detector system.
		The other output is triggered to $k$ clicks of the detector.
	}\label{Fig:AdditionScheme}
	\end{figure}

	In Fig.~\ref{Fig:AdditionScheme}, a scenario for multi-photon addition is outlined.
	An incoming signal is combined with an externally pumped parametric process, which generates two output beams.
	In particular, blocking the signal, a pump photon will generate two output photons propagating in different directions. 
	If one of the photons is measured with an on-off detector system in this case, the corresponding twin photon is 
	propagating in the other spatial mode. In the general case of Fig.~\ref{Fig:AdditionScheme}, the signal field undergoes a parametric amplification process 
	and we post-select the output states to $k$ clicks of the detector.

	First, we consider a coherent input field $|\alpha\rangle$ and the operator $\hat M(\lambda)$ only.
	After the treatment of this special case, we generalize our result to arbitrary states and click POVM elements.
	The considered mixing process is described by the squeezing transformation~\cite{BookVogel,BookAgarwal}
	\begin{align}
		\hat S={\rm e}^{\xi\hat a^\dagger\hat b^\dagger-\xi\hat a\hat b},
	\end{align}
	where we can assume that $\xi$ is a real and positive number.
	The input fields are transformed as
	\begin{align}
		\nonumber
		\hat S\hat a\hat S^\dagger=&\mu\hat a-\nu\hat b^\dagger
		\quad\text{and}\quad
		\hat S\hat b\hat S^\dagger=\mu\hat b-\nu\hat a^\dagger,\\
		\hat S|0,0\rangle=&\frac{1}{\mu}\sum_{m=0}^\infty \left(\frac{\nu}{\mu}\right)^m|m,m\rangle
		=\frac{1}{\mu}{\rm e}^{\frac{\nu}{\mu}\hat a^\dagger\hat b^\dagger}|0,0\rangle,
	\end{align}
	with $\mu=\cosh\xi$, $\nu=\sinh\xi$, and $\mu^2-\nu^2=1$.
	The coherent input state may be written as
	\begin{align}
		|\alpha,0\rangle=\hat D(\alpha)|0,0\rangle={\rm e}^{-\frac{|\alpha|^2}{2}}{\rm e}^{\alpha\hat a^\dagger}|0,0\rangle.
	\end{align}
	This yields an output state in the form:
	\begin{align}
		|\alpha,0\rangle\mapsto|\psi\rangle=\frac{{\rm e}^{-\frac{|\alpha|^2}{2}}}{\mu}{\rm e}^{\alpha[\mu\hat a^\dagger-\nu\hat b]}{\rm e}^{\frac{\nu}{\mu}\hat a^\dagger\hat b^\dagger}|0,0\rangle.
	\end{align}
	Here it is useful to apply the well-known Baker-Campbell-Hausdorff formula: ${\rm e}^{\hat x}{\rm e}^{\hat y}={\rm e}^{[\hat x,\hat y]}{\rm e}^{\hat y}{\rm e}^{\hat x}$, which is valid if $[[\hat x,\hat y],\hat x]=0$ and $[[\hat x,\hat y],\hat y]=0$.
	Hence, for the choice $\hat x=-\nu\alpha\hat b$, $\hat y=\frac{\nu}{\mu}\hat a^\dagger\hat b^\dagger$, and $[\hat x,\hat y]=-\frac{\nu^2}{\mu}\alpha\hat a^\dagger$, the output state can be written as
	\begin{align}
		\nonumber |\psi\rangle=&\frac{{\rm e}^{-\frac{|\alpha|^2}{2}}}{\mu}{\rm e}^{\alpha\mu\hat a^\dagger-\frac{\nu^2}{\mu}\alpha\hat a^\dagger}
		{\rm e}^{\frac{\nu}{\mu}\hat a^\dagger\hat b^\dagger}{\rm e}^{-\nu\alpha\hat b}|0,0\rangle
		\\=&\frac{{\rm e}^{-\frac{|\alpha|^2}{2}}}{\mu}{\rm e}^{\frac{1}{\mu}\alpha\hat a^\dagger}\sum_{m=0}^\infty \left(\frac{\nu}{\mu}\right)^m|m,m\rangle.
	\end{align}
	Now, we perform a partial trace of the second mode with $\hat M(\lambda)$, cf. Eq.~\eqref{Eq:SimpleMeasure}.
	The resulting output can be computed as
	\begin{align}
		\nonumber &{\rm tr}_B\left(|\psi\rangle\langle\psi|\left[\hat 1\otimes {:}{\rm e}^{-\lambda\hat b^\dagger\hat b}{:}\right]\right)
		\\\nonumber =&\frac{{\rm e}^{-|\alpha|^2}}{\mu^2}{\rm e}^{\frac{1}{\mu}\alpha\hat a^\dagger}\left[\sum_{m=0}^\infty\left((1-\lambda)\frac{\nu^2}{\mu^2}\right)|m\rangle\langle m|\right]{\rm e}^{\frac{1}{\mu}\alpha^\ast\hat a}
		\\\nonumber =&\frac{{\rm e}^{-|\alpha|^2}}{\mu^2}{\rm e}^{\frac{1}{\mu}\alpha\hat a^\dagger}{:}{\rm e}^{\left[(1-\lambda)\frac{\nu^2}{\mu^2}-1\right]\hat a^\dagger\hat a}{:}{\rm e}^{\frac{1}{\mu}\alpha^\ast\hat a}
		\\ =&\frac{1}{\mu^2}{:}{\rm e}^{-\frac{1}{\mu^2}(\hat a-\mu\alpha)^\dagger(\hat a-\mu\alpha)}{\rm e}^{-\lambda\frac{\nu^2}{\mu^2}\hat a^\dagger\hat a}{:}
		\label{Eq:AddNO}
		\\\nonumber =&\sum_{m=0}^\infty\frac{1}{m!}\left(-\lambda\frac{\nu^2}{\mu^2}\right)^m\hat a^\dagger{}^m{:}\frac{1}{\mu^2}{\rm e}^{-\frac{1}{\mu^2}(\hat a-\mu\alpha)^\dagger(\hat a-\mu\alpha)}{:}\hat a{}^m.
	\end{align}
	For further considerations, it is useful to have a closer look at the normally ordered term.
	It turns out that this expression represents a displaced thermal state
	\begin{align}
		\nonumber &\Lambda^{\rm (noise)}_{\mu}(|\alpha\rangle\langle \alpha|)={:}\frac{1}{\mu^2}{\rm e}^{-\frac{1}{\mu^2}(\hat a-\mu\alpha)^\dagger(\hat a-\mu\alpha)}{:}
		\\\nonumber=&\hat D(\mu\alpha)\left[\frac{1}{\mu^2}\sum_{n=0}^\infty \left(\frac{\mu^2-1}{\mu^2}\right)^n|n\rangle\langle n|\right]\hat D(\mu\alpha)^\dagger
		\\=&\int{\rm d}^2\alpha'\,\frac{1}{\pi (\mu^2-1)}{\rm e}^{-\frac{|\alpha'-\mu\alpha|^2}{\mu^2-1}}|\alpha'\rangle\langle\alpha'|.
	\end{align}
	Hence, the $P$~function of a general input state has to be convoluted with this noise prior to the addition process itself,
	\begin{align}
		P^{\rm (noise)}_{\mu}(\alpha')=\int\!{\rm d}^2\alpha\,\frac{{\rm e}^{-\frac{|\alpha'-\mu\alpha|^2}{\mu^2-1}}}{\pi (\mu^2-1)}P(\alpha).
	\end{align}
	Is is also worth mentioning that the coherent amplitude of the input state is amplified by the factor $\mu$.

	For the particular measurement of $\hat\Pi_k$ in the form of Eq.~\eqref{Eq:ClickPOVM}, we get the output state as
	\begin{align}
		\nonumber \hat\rho_{k,\rm out}=&\sum_{m=0}^\infty \frac{1}{m!}\left[\binom{N}{k}\sum_{j=1}^k\binom{k}{j}(-1)^{k-j}\right.
		\\\nonumber &\left.\times\left(-\left[\eta-\eta\frac{j}{N}\right]\frac{\nu^2}{\mu^2}\right)^m\right] \hat a^\dagger{}^m\Lambda_{\mu}^{\rm (noise)}(\hat\rho_{\rm in})\hat a{}^m
		\\=&\sum_{m=k}^\infty \frac{1}{m!}\mathcal D_{k,m}^{-\eta',\eta'}\hat a^\dagger{}^m\Lambda^{\rm (noise)}_{\mu}(\hat\rho_{\rm in})\hat a{}^m,
		\label{Eq:AdditionKrauss}
	\end{align}
	with a resulting efficiency of
	\begin{align}
		\eta'=\eta\frac{\nu^2}{\mu^2}.
	\end{align}
	The general input-output relation in Eq.~\eqref{Eq:AdditionKrauss} is the quantum description of the process in Fig.~\ref{Fig:AdditionScheme}.
	Again, the output state is normalized to the probability ${\rm tr}\,\hat\rho_{k,\rm out}$ to have $k$ clicks at the detector system.
	The attenuation given by $\Lambda^{\rm (noise)}_\mu$ is due to a mixing of the input field with vacuum in the non-linear medium, and not the result of detector imperfections.
	Let us also mention the similarity to the subtraction process.
	The input-output relations in Eqs.~\eqref{Eq:SubtractionKrauss} and~\eqref{Eq:AdditionKrauss} share the same formal structure, except for an exchange of annihilation and creation operators as well as an exchange of a loss process with an amplification introducing thermal noise.
	Moreover, even the no-click count event can be of some interest, cf.~\cite{ASW93}.
	For $k=0$ the process in Eq.~\eqref{Eq:AdditionKrauss} maps a coherent input -- up to a normalization constant -- to a displaced thermal state, see Appendix~\ref{App:DThS}.
	From the variance $\sigma^2$ of the thermalized output $P$~function, one can directly compute the squeezing parameter $\xi$,
	\begin{align}
		\sigma^2=\frac{\nu^2(1-\eta)}{1+\eta\nu^2}
		\quad\text{and}\quad \nu=\sinh\xi.
	\end{align}

\begin{figure}[ht]
	\includegraphics*[width=8.4cm]{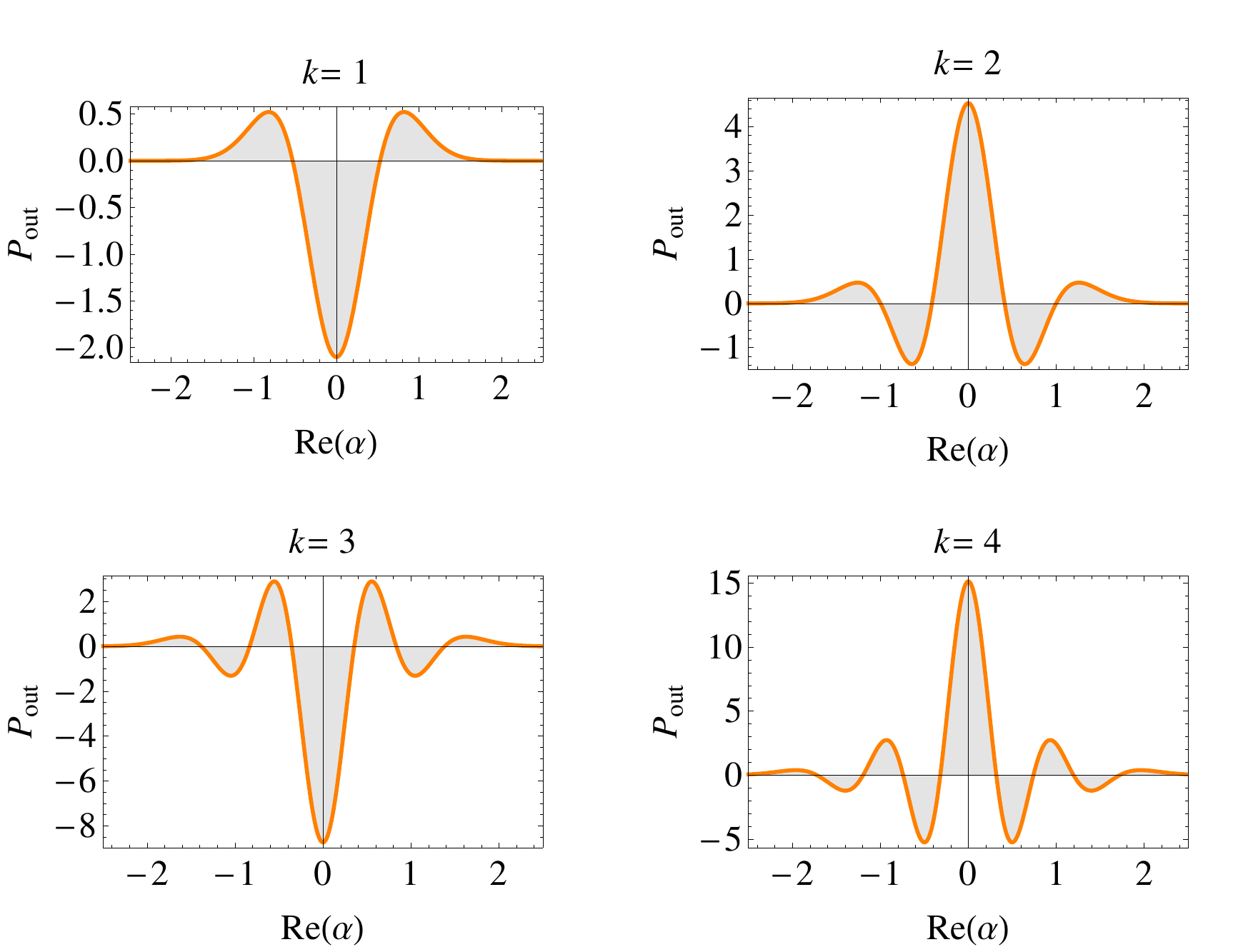}
	\caption{
		(Color online)
		The plots show the $P$~function of a $k$-click conditioned thermal state generated by a click counting detector with $N=16$ on-off diodes and a quantum efficiency $\eta=0.8$.
		The considered squeezing corresponds to $\mu=\cosh\xi=1.4$.
		The mean thermal photon number of the input state is $\bar n=0.5$.
	}\label{Fig:AdditionExample}
	\end{figure}

	As an example for the addition protocol under study, we will show the output for the thermal input state in Eq.~\eqref{Eq:ThermalState}.
	The noise convolution yields -- as a convolution of two Gaussian functions -- an effective noise contribution of $\bar n_0=\mu^2(\bar n+1)-1$.
	The $k$ click conditioned state in Eq.~\eqref{Eq:AdditionKrauss} may be written in normally ordered form as
	\begin{align}
		\nonumber
		\hat\rho_{\rm out}=&{:}\binom{N}{k}
		\left({\rm e}^{-\eta'\hat n/N}\right)^{N-k}
		\\&\times
		\left(\hat 1-{\rm e}^{-\eta'\hat n/N}\right)^{k}
		\frac{{\rm e}^{-\hat n/(\bar n_0+1)}}{\bar n_0+1}{:},
	\end{align}
	see also Eq.~\eqref{Eq:AddNO} or Appendix~\ref{App:DThS}, where the more general example of a displaced thermal state is presented.
	The $P$~function can be easily extracted from an expansion of the $k$th power and the relation ($\lambda>0$):
	\begin{align}
		{:}\frac{{\rm e}^{-\frac{1}{\lambda+1}\hat a^\dagger\hat a}}{\lambda+1}{:}
		=\int\!{\rm d}^2\alpha\,\frac{{\rm e}^{-\frac{1}{\lambda}|\alpha|^2}}{\pi\lambda}|\alpha\rangle\langle\alpha|.
	\end{align}
	For different numbers of clicks, $k$, we observe different numbers of oscillations within the output $P$~function between negative and positive values (Fig.~\ref{Fig:AdditionExample}).
	Hence, we directly verify the nonclassical features of the considered process.
	The quantum correlated output states introduce a new class of nonclassical states with a regular $P$~function.
	Similar to the prominent single-/multi-photon added thermal states~\cite{AT92,KVPZB08}, these states may be denoted as $k$-click conditioned thermal states.

\section{Beyond Noiseless Amplification}\label{Sec:NA}

	So far, we studied processes conditioned to a measurement of $k$ clicks with systems of avalanche diodes.
	Now, we can study a manifold of combinations of these individual procedures.
	A prominent example is an operation which applies one addition followed by a subtraction:
	\begin{align}\label{Eq:NAtypically}
		\hat\rho_{\rm in}\mapsto\frac{\hat a\hat a^\dagger\hat\rho_{\rm in}\hat a\hat a^\dagger}{{\rm tr}\left[\hat a\hat a^\dagger\hat\rho_{\rm in}\hat a\hat a^\dagger\right]}.
	\end{align}
	It represents a noiseless amplification protocol as it has been experimentally realized in Ref.~\cite{ZFB11}.
	In the limit of small coherent amplitudes, $|\beta|\ll1$, this process acts like 
	\begin{align}
		|\beta\rangle\approx |0\rangle+\beta|1\rangle\mapsto|0\rangle+2\beta|1\rangle\approx |2\beta\rangle,
	\end{align}
	i.e., we have a gain of two in the coherent amplitude without the addition of noise.
	For a recent study on quantum limits of amplification protocols, we refer to Ref.~\cite{PJCC13}.

\begin{figure}[ht]
	\includegraphics*[width=8.4cm]{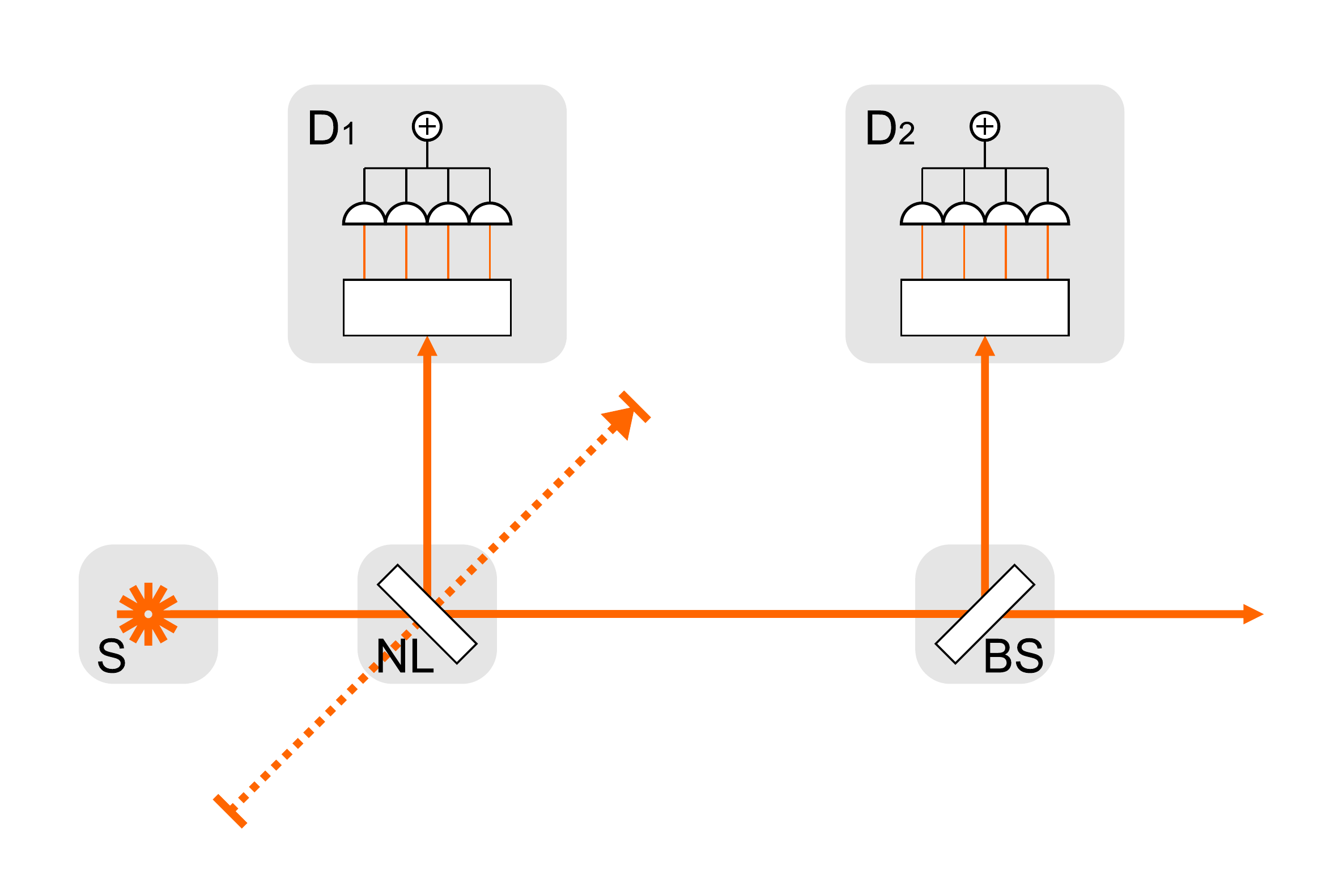}
	\caption{
		(Color online)
		A combination of an initial addition followed by the subtraction of photons with click counting detectors is shown.
		The first detector D${}_1$ consists of $N_1$ diodes with a quantum efficiency of $\eta_1$.
		The second detector D${}_2$ consists of $N_2$ diodes with a quantum efficiency of $\eta_2$.
		The final outcome is conditioned to $k_1$ clicks of the first and $k_2$ clicks of the second detector.
	}\label{Fig:NoiselessAmplificationScheme}
	\end{figure}

	Let us consider the generalization of this process to a combination of a $k_1$-photon addition followed by a $k_2$-photon subtraction.
	Applying click counting detectors means that the scheme in Fig.~\ref{Fig:AdditionScheme} is combined with the setup in Fig.~\ref{Fig:SubtractionScheme}.
	This leads to the process depicted in Fig.~\ref{Fig:NoiselessAmplificationScheme}.
	We are going to study the output states which correspond to $k_1$ clicks of the first and $k_2$ clicks of the second detector.

	\begin{figure*}[ht!]
	\includegraphics*[width=17.6cm]{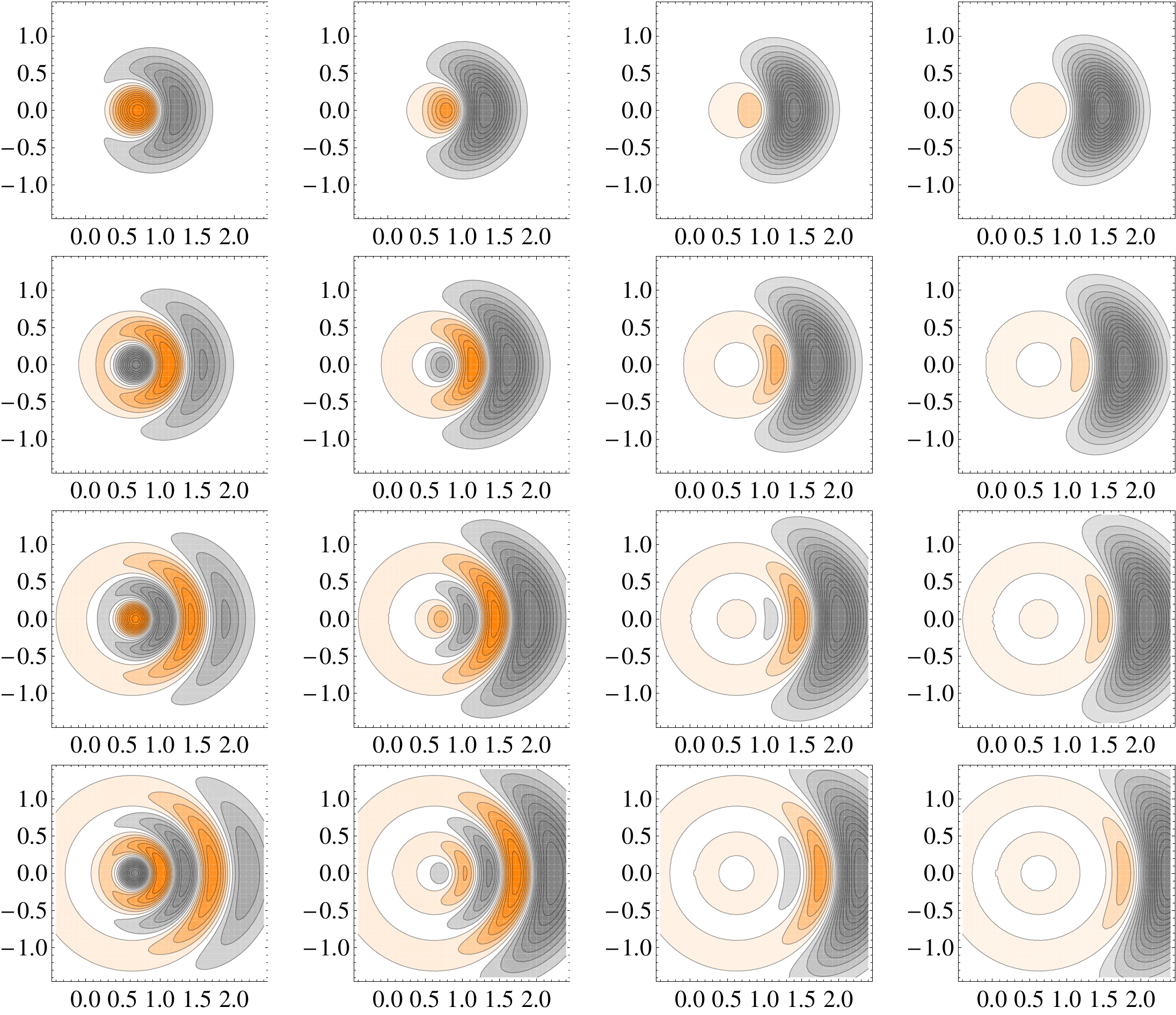}
	\caption{
		(Color online)
		Contour plot of the output $P$~function for a coherent input state with $\beta=1/\sqrt{2}$.
		Both detectors consist of the same number of diodes $N_1=N_2=4$ with identical quantum efficiencies of only $\eta_1=\eta_2=0.5$.
		The parametric process is described by $\mu=3/2$ and for the beam splitter holds $t=2/3$.
		The number of the row is equal to the number of additive clicks, $k_1=1,2,3,4$, whereas the column counts the subtractions, $k_2=0,1,2,3$.
		A dark orange color represents a highly negative contribution, whereas gray depicts positive parts of the $P$~function.
	}\label{Fig:NoiselessAmplificationExample}
	\end{figure*}

	Let us study the scenario of quantum state engineering based on the full process description, as a generalization of the noiseless amplification process.
	In Secs.~\ref{Sec:Addition} and~\ref{Sec:Subtraction}, we gave the corresponding input-output equations of the individual processes.
	Hence, we obtain after some algebra that a coherent input state $|\beta\rangle$ is mapped to the output $P$~function:
	\begin{align}\label{Eq:NAout}
		&P_{(k_1,k_2),\rm out}(\alpha;\beta)
		\\=&\sum_{j_1=0}^{k_1}\sum_{j_2=0}^{k_2} \frac{f_{j_1,j_2}}{\pi}
		{\rm e}^{{-}\lambda_{2;j_1,j_2}|\alpha|^2{+}2\lambda_{1;j_1,j_2}{\rm Re}(\alpha\beta^\ast){-}\lambda_{0;j_1,j_2}|\beta|^2}.\nonumber
	\end{align}
	which is normalized to the probability that $k_1$ addition and $k_2$ absorption processes have been realized simultaneously.
	The occurring coefficients are
	\begin{align}
		\nonumber f_{j_1,j_2}{=}&\binom{N_1}{k_1}\!\binom{N_2}{k_2}\!\binom{k_1}{j_1}\!\binom{k_2}{j_2}\!\frac{(-1)^{k_1-j_1+k_2-j_2}}{t^2\nu^2(1-\eta_1[1{-}j_1/N_1])},
		\\\nonumber \lambda_{2;j_1,j_2}{=}&\frac{1+\eta_1\nu^2[1{-}j_1/N_1]}{t^2\nu^2(1-\eta_1[1{-}j_1/N_1])}+\frac{\eta_2 r^2[1{-}j_2/N_2]}{t^2},
		\\\nonumber \lambda_{1;j_1,j_2}{=}&\frac{\mu}{t\nu^2(1-\eta_1[1{-}j_1/N_1])},
		\\\lambda_{0;j_1,j_2}{=}&1+\frac{1}{\nu^2(1-\eta_1[1{-}j_1/N_1])}.
	\end{align}
	We therefore get for a general input state, $P_{\rm in}(\beta)$, the output state as
	\begin{align}
		\nonumber \hat\rho_{(k_1,k_2),\rm out}=&\int\!{\rm d}^2\alpha\left[\int\!{\rm d}^2\beta\,P_{\rm in}(\beta)\right.
		\\ &\left.\phantom{\int}\times P_{(k_1,k_2),\rm out}(\alpha;\beta)\right]|\alpha\rangle\langle\alpha|.
	\end{align}
	Note that some terms in Eq.~\eqref{Eq:NAout} can lead to $\delta$-shaped contributions, if the denominator of the coefficients become zero.
	However, this only occurs in the unphysical case $\eta_1\to100\%$ (for $\nu,t\neq0$).

	In Fig.~\ref{Fig:NoiselessAmplificationExample}, the output $P$~function for a coherent input state is given.
	The first row corresponds to $k_1=1$ click and the last row to $k_1=4$ clicks of the click addition part.
	The first column represents $k_2=0$ clicks and the last column $k_2=3$ clicks of the subtraction.
	It can be directly observed that all these engineered phase-space distributions exhibit nonclassical features (orange, negative contributions).
	The higher the number of clicks $k_1$ the more negative interference fringes appear.
	In contrast to the nonclassical click addition, the subtraction diminishes these quantum features together with a deformation of the positive (gray) contributions. 
	The given examples clearly show that the considered setup renders it possible to prepare various nonclassical states with different types of quantum interference effects.
 	The probabilities for the realization of the individual click combinations are listed in Table~\ref{Tab:NoiselessAmplificationExample}.

	\begin{table}
		\caption{
			Probability for the realization of a $(k_1,k_2)$ conditioned output state.
			The rows are numbered by the addition clicks, $k_1$, and the columns by the subtraction clicks, $k_2$.
			The measurement is done for the same parameters as in Fig.~\ref{Fig:NoiselessAmplificationExample}.
		}\label{Tab:NoiselessAmplificationExample}
		\begin{tabular}{|c|ccccc|}
			\hline
			${\rm tr}\,\hat\rho_{(k_1,k_2),\rm out}$ & 0 & 1 & 2 & 3 & 4 \\
			\hline
			0 & \,\,16.80\%\,\, & 8.83\% & \,\,2.47\%\,\, & \,\,0.39\%\,\, & \,\,0.03\%\,\, \\
			1 & 8.46\% & \,\,12.38\%\,\, & 6.85\% & 1.88\% & 0.22\% \\
			2 & 3.17\% & 8.24\% & 7.90\% & 3.54\% & 0.65\% \\
			3 & 0.81\% & 3.32\% & 4.99\% & 3.48\% & 0.99\% \\
			4 & 0.11\% & 0.67\% & 1.52\% & 1.60\% & 0.70\% \\
			\hline
		\end{tabular}
	\end{table}

	Our general theoretical treatment enables experimentalists to predict and generate various nonclassical states in arbitrary ranges of parameters.
	As shown, the same experimental setup can lead to many different forms of nonclassical correlations even with asymmetric phase-space distributions.
	For an easily accessible, coherent input state the process led to quantum states with a regular $P$~function, contrary to the idealized noiseless amplification process in Eq.~\eqref{Eq:NAtypically}.
	This allows a direct sampling of the $P$ function from experimental data which have been measured, for example, by balanced homodyne detection, cf.~\cite{KVPZB08}.
	Let us also note that the plots in Fig.~\ref{Fig:NoiselessAmplificationExample} have been obtained for a quantum efficiency of only $50\%$.
	In the case of the photoelectric counting theory, the low efficiency domain can be used for the realization of quantum-mechanical weak values of observables~\cite{AP07}.

	Finally, let us outline possible generalizations of the presented schemes.
	First, various combinations of heralding, addition, and subtraction scenarios with click counting detectors could be combined.
	Second, one can condition the output not to single click events, but to multiple ones.
	The resulting state is a mixing of the individual output states for different numbers of clicks.
	Similarly, one can withdraw only a certain percentage of states which are realized.
	This would allow one to steer the mixing ratio between different numbers of clicks.
	Third,	at one input port of the addition and subtraction protocols vacuum was considered.
	A straight-forward extension of the presented approach could use this free port, e.g., for ancilla states as they are applied in some amplification scenarios.
	Fourth, additional features may appear when generalized detection processes are involved, e.g., two-photon absorption within the diodes, cf.~\cite{SVAPRA13}.
	Similarly, arbitrary noise models may be included at any stage of the protocol description.
	Moreover, a propagation in non-linear media, e.g., in fibers, or higher order wave mixing may further enhance the applicability of click detector systems for quantum state engineering.


\section{Conclusions}\label{Sec:Conclusions}

	In conclusion, we have derived analytical input-output relations for the application of on-off detector systems in quantum state engineering.
	This includes the use of a single avalanche photo diode up to any number of diodes in detector arrays or multiplexing schemes.
	Based on the click counting statistics of such detector systems, we could derive a proper description of several protocols and arbitrary input states.
	First, we considered the heralded photon generation describing the fundamentals of our approach, e.g., the positive operator valued measure for click counting.
	Second, we studied two prominent processes: photon addition and photon subtraction.
	The click triggered outcome of these processes have been derived for arbitrary input states in terms of input-output relations.

	As explicit examples, we studied displaced thermal states serving as an input.
	This led to another class of click-conditioned displaced thermal states.
	As a combination of addition and subtraction, we considered a noiseless amplification protocol.
	We showed for this example a manifold of new quantum states having nonclassical features which can be directly generated by a post-selection to particular click numbers.
	Additionally, we computed the probabilities for the occurrence of the desired output states of these non-deterministic processes.
	Throughout the present work, we consistently included the quantum efficiency of the detection process.
	The occurring efficiencies of the diodes have been studied in high and low efficiency regimes.
	The approach also revealed unavoidable attenuation of the process which has to be taken into account.
	Our results present a useful scheme for the choice of experimental conditions in order to realize tailor-made nonclassical states with systems of on-off detectors.


\section*{Acknowledgments}
	This work was supported by the Deutsche Forschungsgemeinschaft through SFB 652.
	J.S. gratefully acknowledges financial support from the Oklahoma State University.

\appendix


\section{${\mathcal D}$-symbol}\label{App:Dsym}

	The symbol ${\mathcal D}^{\tau,\sigma}_{k,m}$ is defined as
	\begin{align*}
		{\mathcal D}^{\tau,\sigma}_{k,m}=\binom{N}{k}\lim_{x\to0}\partial_x^m\left[{\rm e}^{\tau x}\left({\rm e}^{\frac{\sigma}{N}x}-1\right)^{k}\right].
	\end{align*}
	Expanding the $k$th power, we get the representation:
	\begin{align*}
		{\mathcal D}^{\tau,\sigma}_{k,m}=&\binom{N}{k}\sum_{j=0}^k\binom{k}{j}(-1)^{k-j}\lim_{x\to0}\partial_x^m{\rm e}^{\left(\tau+\frac{\sigma}{N}j\right)x}
		\\=&\binom{N}{k}\sum_{j=0}^k\binom{k}{j}(-1)^{k-j}\left(\tau+\frac{\sigma}{N}j\right)^m.
	\end{align*}
	From the computational point of view, a recursion relation is a practicable tool.
	The initial values can be readily obtained as
	\begin{align*}
		{\mathcal D}^{\tau,\sigma}_{0,0}=&1,\\
		{\mathcal D}^{\tau,\sigma}_{k,0}=&0 \,\,\,\,\, \text{ for } \, k>0,\\
		\text{and } {\mathcal D}^{\tau,\sigma}_{0,m}=&\tau^m \text{ for } m>0.
	\end{align*}
	The recursion relation is derived as
	\begin{align*}
		{\mathcal D}^{\tau,\sigma}_{k,m}=&\binom{N}{k}\lim_{x\to0}\partial_x^{m-1}\left[{\rm e}^{\tau x}\left({\rm e}^{\frac{\sigma}{N}x}-1\right)^{k}\tau\right.
		\\&\left.+{\rm e}^{\tau x}\left({\rm e}^{\frac{\sigma}{N}x}-1\right)^{k-1}k\left({\rm e}^{\frac{\sigma}{N}x}-1+1\right)\frac{\sigma}{N}\right]
		\\=&\left[\tau+\sigma\frac{k}{N}\right]{\mathcal D}^{\tau,\sigma}_{k,m{-}1}+\sigma\frac{N-k+1}{N}{\mathcal D}^{\tau,\sigma}_{k{-}1,m{-}1}.
	\end{align*}
	Note that this relation and the initial values imply that
	\begin{align*}
		{\mathcal D}^{\tau,\sigma}_{k,m}=0 \text{ for } k>m.
	\end{align*}
	Using the definition $\binom{N}{k}=0$ for $k>N$, we additionally get ${\mathcal D}^{\tau,\sigma}_{k,m}=0$ for $k>N$.	
	Moreover, the recursion yields the diagonal elements as
	\begin{align*}
		&{\mathcal D}^{\tau,\sigma}_{k,k}=\sigma\frac{N-k+1}{N}{\mathcal D}^{\tau,\sigma}_{k{-}1,k{-}1}
		=\ldots = \frac{\sigma^k}{N^k}\frac{N!}{(N-k)!}.
	\end{align*}


\section{Error estimation}\label{App:ErrE}

	We are going to derive an upper bound to the error one would obtain if a Poissonian measurement is applied instead of the binomial one.
	The standard photoelectric counting theory is based on the POVM elements
	\begin{align*}
		\hat P_k=&{:}\frac{(\eta\hat n)^k}{k!}{\rm e}^{-\eta\hat n}{:}=\frac{\eta^k}{k!}{:}\hat n^k{\rm e}^{(1-\eta)\hat n}{\rm e}^{-\hat n}{:}\nonumber\\
		=&\sum_{m=k}^\infty \binom{m}{k}\eta^k(1-\eta)^{m-k}|m\rangle\langle m|,
	\end{align*}
	for $k=0,1,2,\ldots$ and the quantum efficiency $\eta$.
	The deviation between the expectation value ${\rm tr}(\hat\rho\hat\Pi_k)$ for $k$ clicks, cf. Eq.~\eqref{Eq:DsymClickPOVM}, and ${\rm tr}(\hat\rho\hat P_k)$ is
	\begin{align*}
		|{\rm tr}(\hat \rho[\hat P_k-\hat\Pi_k])|\leq \|\hat\rho\|_{\rm tr} \|\hat P_k-\hat\Pi_k\|_{\rm Op},
	\end{align*}
	where we applied H\"older's inequality.
	Since the trace norm is $\|\hat\rho\|_{\rm tr}=1$, we get an upper bound of the deviation by the operator norm $\|\hat P_k-\hat\Pi_k\|_{\rm Op}$.
	This norm can be calculated by
	\begin{align*}
		\|\hat P_k-\hat\Pi_k\|_{\rm Op}=\sup_{m\geq k}\left|\binom{m}{k}\eta^k(1-\eta)^{m-k}-\mathcal D^{1-\eta,\eta}_{k,m} \right|,
	\end{align*}
	because both, $\hat P_k$ and $\hat\Pi_k$, are diagonal in a common basis.
	Note that $\hat\Pi_k\to\hat P_k$ for $N\to\infty$, see~\cite{SVAPRA12} for the corresponding statistics, which implies that $\|\hat P_k-\hat\Pi_k\|_{\rm Op}\to0$.
	For $\eta\to 1$, we would additionally get the $k$-photon projector: $\hat P_k\to |k\rangle\langle k|$.


\section{Click conditioned displaced thermal states}\label{App:DThS}

	We consider -- as input states for the addition and subtraction -- the class of displaced thermal states, which can be given by some equivalent representations:
	\begin{align*}
		\hat\rho_{\rm in}=&\int\! {\rm d}^2\alpha\,\frac{{\rm e}^{-|\alpha-\alpha_0|^2/\bar n}}{\pi\bar n}\,|\alpha\rangle\langle\alpha|
		\\=&\frac{1}{\bar n+1}\hat D(\alpha_0)\left[\sum_{n=0}^\infty\left(\frac{\bar n}{\bar n+1}\right)^n|n\rangle\langle n|\right]\hat D(\alpha_0)^\dagger
		\\=&\frac{1}{\bar n+1}{:}{\rm e}^{-(\hat a-\alpha_0)^\dagger(\hat a-\alpha_0)/(\bar n+1)}{:},
	\end{align*}
	where $\alpha_0$ denotes the displacement, $\bar n$ denotes the mean thermal photon number (for zero displacement), and $\hat D(\alpha_0)$ is the displacement operator.
	Note that $\bar n\to0$ yields the coherent state $|\alpha_0\rangle\langle\alpha_0|$.

	Following the multi-photon subtraction protocol, we observe that the state $|\alpha\rangle$ is mapped to
	\begin{align*}
		 &\binom{N}{k}\sum_{j=0}^k\binom{k}{j}(-1)^{k-j}{\rm e}^{-\eta\left(1-\frac{j}{N}\right)r^2|\alpha|^2}|t\alpha\rangle\langle t\alpha|
		\\=&\binom{N}{k}\left({\rm e}^{-\frac{\eta r^2|\alpha|^2}{N}}\right)^{N-k}\left(1-{\rm e}^{-\frac{\eta r^2|\alpha|^2}{N}}\right)^{k}|t\alpha\rangle\langle t\alpha|
	\end{align*}
	Hence, we obtain for the displaced thermal state the output state of a $k$ click subtraction as
	\begin{align*}
		\hat\rho_{k,\rm out}=\int\! {\rm d}^2\alpha\,&\left[\frac{1}{\pi t^2\bar n}\binom{N}{k}{\rm e}^{-\frac{|\alpha-t\alpha_0|^2}{t^2\bar n}}\left({\rm e}^{-\frac{\eta r^2|\alpha|^2}{t^2 N}}\right)^{N-k}\right.
		\\&\left.\times\left(1-{\rm e}^{-\frac{\eta r^2|\alpha|^2}{t^2N}}\right)^{k}\right]|\alpha\rangle\langle \alpha|,
	\end{align*}
	including the substitution $t\alpha \mapsto \alpha$.
	The probability for obtaining this output state is
	\begin{align*}
		{\rm tr}\,\hat\rho_{k,\rm out}=&\binom{N}{k}\sum_{j=0}^k\binom{k}{j}(-1)^{k-j}\frac{1}{\gamma_j}{\rm e}^{-\left(1-\frac{1}{\gamma_j}\right)\frac{|\alpha_0|^2}{\bar n}},
		\\\text{with }
		\gamma_j=&1+\eta r^2\bar n\left(1-\frac{j}{N}\right).
	\end{align*}

	Let us now focus on the addition protocol.
	We conclude from Eq.~\eqref{Eq:AddNO} that the state $|\alpha\rangle$ is mapped to
	\begin{align*}
		&\binom{N}{k}\sum_{j=0}^k\binom{k}{j}(-1)^{k-j}\frac{1}{\mu^2}
		\\&\times{:}{\rm e}^{-\frac{1}{\mu^2}(\hat a-\mu\alpha)^\dagger(\hat a-\mu\alpha)}{\rm e}^{-\eta\left(1-\frac{j}{N}\right)\frac{\nu^2}{\mu^2}\hat a^\dagger\hat a}{:}
		\\=&{:}\binom{N}{k}\left({\rm e}^{-\frac{\eta\nu^2}{\mu^2 N}\hat a^\dagger\hat a}\right)^{N-k}\left(\hat 1-{\rm e}^{-\frac{\eta\nu^2}{\mu^2 N}\hat a^\dagger\hat a}\right)^{k}
		\\ &\times\frac{1}{\mu^2}{\rm e}^{-\frac{1}{\mu^2}(\hat a-\mu\alpha)^\dagger(\hat a-\mu\alpha)}{:}\,.
	\end{align*}
	Later on, we give the analytical $P$~function for such a normally ordered expression.
	Convolving this expression with the displaced thermal input, we get
	\begin{align*}
		\hat\rho_{k,\rm out}=&{:}\binom{N}{k}\left({\rm e}^{-\frac{\eta\nu^2}{\mu^2 N}\hat a^\dagger\hat a}\right)^{N-k}\left(\hat 1-{\rm e}^{-\frac{\eta\nu^2}{\mu^2 N}\hat a^\dagger\hat a}\right)^{k}
		\\ &\times\frac{1}{\mu^2(\bar n+1)}{\rm e}^{-\frac{1}{\mu^2(\bar n+1)}(\hat a-\mu\alpha_0)^\dagger(\hat a-\mu\alpha_0)}{:}\,.
	\end{align*}
	The normalization is
	\begin{align*}
		{\rm tr}\,\hat\rho_{k,\rm out}=&\binom{N}{k}\sum_{j=0}^k\binom{k}{j}(-1)^{k-j}\frac{1}{\gamma_j}{\rm e}^{-\left(1-\frac{1}{\gamma_j}\right)\frac{|\alpha_0|^2}{\bar n+1}},
		\\\text{with }
		\gamma_j=&1+\eta \nu^2(\bar n+1)\left(1-\frac{j}{N}\right).
	\end{align*}

	For the computation of the results, we used the following simple relations:
	the relation between the normal ordered representation and the $P$~function of thermal states ($0<\lambda_2<1$),
	\begin{align*}
		&{:}{\rm e}^{-\lambda_2\hat a^\dagger\hat a+\lambda_1\hat a^\dagger+\lambda_1^\ast\hat a}{:}
		\\=&\int\!{\rm d}^2\alpha\,\frac{{\rm e}^{-(\lambda_2|\alpha|^2-\lambda_1\alpha^\ast-\lambda_1^\ast\alpha+|\lambda_1|^2)/(1-\lambda_2)}}{\pi(1-\lambda_2)}|\alpha\rangle\langle\alpha|,
	\end{align*}
	the Gaussian integral formula ($\lambda_2>0$),
	\begin{align*}
		\int\!{\rm d}^2\alpha\,{\rm e}^{-\lambda_2|\alpha|^2+\lambda_1^\ast\alpha+\lambda_1\alpha^\ast}=\frac{\pi}{\lambda_2}{\rm e}^{|\lambda_1|^2/\lambda_2},
	\end{align*}
	and -- as a combination of the previous formulas -- the following trace ($0<\lambda_2<1$):
	\begin{align*}
		{\rm tr}\,{:}{\rm e}^{-\lambda_2\hat a^\dagger\hat a+\lambda_1\hat a^\dagger+\lambda_1^\ast\hat a}{:}=\frac{{\rm e}^{|\lambda_1|^2/\lambda_2}}{\lambda_2}.
	\end{align*}
	It is also useful to recall the displacement operation
	$\hat D(\lambda_1){:}{\rm e}^{-\hat a^\dagger\hat a} {:}\hat D(\lambda_1)^\dagger$=${:}{\rm e}^{-(\hat a-\lambda_1)^\dagger(\hat a-\lambda_2)} {:}$
	and the representation ${:}{\rm e}^{-\lambda_2\hat a^\dagger\hat a} {:}$=$\sum_{m=0}^\infty(1-\lambda_2)^m |m\rangle\langle m|$.



\begin{thebibliography}{99}
	\bibitem{KMNRDM07} P. Kok, W. J. Munro, K. Nemoto, T. C. Ralph, J. P. Dowling, and G. J. Milburn, Rev. Mod. Phys. {\bf 79}, 135 (2007).
	\bibitem{HM86} C. K. Hong and L. Mandel, Phys. Rev. Lett. {\bf 56}, 58 (1986).
	\bibitem{JHBL10} N. Jain, S. R. Huisman, E. Bimbard, and A. I. Lvovsky, Opt. Express {\bf 18}, 18254 (2010).
	\bibitem{WLRBGCZCP10} C. Wagenknecht, C.-M. Li, A. Reingruber, X.-H. Bao, A. Goebel, Y.-A. Chen, Q. Zhang, K. Chen, and J.-W. Pan, Nature Photon. {\bf 4}, 549 (2010).
	\bibitem{SCSWS11} C. S\"oller, O. Cohen, B. J. Smith, I. A. Walmsley, and Ch. Silberhorn, Phys. Rev. A {\bf 83}, 031806(R) (2011).
	\bibitem{AMFL13} V. D'Auria, O. Morin, C. Fabre, and J. Laurat, Eur. Phys. J. D {\bf 66}, 249 (2012).
	\bibitem{BDSJBDSW12} T. J. Bartley, G. Donati, J. B. Spring, X.-M. Jin, M. Barbieri, A. Datta, B. J. Smith, and I. A. Walmsley, Phys. Rev. A {\bf 86}, 043820 (2012).
	\bibitem{CXRVHSRKSCE13} M. J. Collins, C. Xiong, I. H. Rey, T. D. Vo, J. He, S. Shahnia, C. Reardon, T. F. Krauss, M. J. Steel, A. S. Clark, and B. J. Eggleton, Nature Commun. {\bf 4}, 2582 (2013).
	\bibitem{FFWSACSLM13} M. F\"ortsch, J. U. F\"urst, C. Wittmann, D. Strekalov, A. Aiello, M. V. Chekhova, Ch. Silberhorn, G. Leuchs, and Ch. Marquardt, Nature Commun. {\bf 4}, 1818 (2013).
	\bibitem{BC10} G. S. Buller and R. J. Collins, Meas. Sci. Technol. {\bf 21}, 012002 (2010).
	\bibitem{SSTT13} H. Shibata, K. Shimizu, H. Takesue, and Y. Tokura, Appl. Phys. Express {\bf 6}, 072801 (2013).
	\bibitem{MVSHLGVBSMN13} F. Marsili, V. B. Verma, J. A. Stern, S. Harrington, A. E. Lita, T. Gerrits, I. Vayshenker, B. Baek, M. D. Shaw, R. P. Mirin, and S. W. Nam, Nature Photon. {\bf 7}, 210 (2013).
	\bibitem{NPGGSBKALSLS14} A. K. Nowak S. L. Portalupi, V. Giesz, O. Gazzano, C. Dal Savio, P.-F. Braun, K. Karrai, C. Arnold, L. Lanco, I. Sagnes, A. Lema\^{\i}tre, and P. Senellart, Nature Commun. {\bf 5}, 3240 (2014).
	\bibitem{ABA10} A. Allevi, M. Bondani, and A. Andreoni, Opt. Lett. {\bf 35}, 1707 (2010).
	\bibitem{DYSTS11} J. F. Dynes, Z. L. Yuan, A. W. Sharpe, O. Thomas, and A. J. Shields, Opt. Express {\bf 19}, 13268 (2011).
	\bibitem{ASSBW03} D. Achilles, Ch. Silberhorn, C. \'Sliwa, K. Banaszek, and I. A. Walmsley, Opt. Lett. {\bf 28}, 2387 (2003).
	\bibitem{FJPF03} M. J. Fitch, B. C. Jacobs, T. B. Pittman, and J. D. Franson, Phys. Rev. A {\bf 68}, 043814 (2003).
	\bibitem{ZABGGBRP05} G. Zambra, A. Andreoni, M. Bondani, M. Gramegna, M. Genovese, G. Brida, A. Rossi, and M. G. A. Paris, Phys. Rev. Lett. {\bf 95}, 063602 (2005).
	\bibitem{FLCEPW09} A. Feito, J. S. Lundeen, H. Coldenstrodt-Ronge, J. Eisert, M. B. Plenio, and I. A. Walmsley, New J. Phys. {\bf 11}, 093038 (2009).
	\bibitem{WDSBY04} E. Waks, E. Diamanti, B. C. Sanders, S. D. Bartlett, and Y. Yamamoto, Phys. Rev. Lett. {\bf 92}, 113602 (2004).
	\bibitem{JDC07} L. A. Jiang, E. A. Dauler, and J. T. Chang, Phys. Rev. A {\bf 75}, 062325 (2007).
	\bibitem{DBIMHCE13} L. Dovrat, M. Bakstein, D. Istrati, E. Megidish, A. Halevy, L. Cohen, and H. S. Eisenberg, Phys. Rev. A {\bf 87}, 053813 (2013).
	\bibitem{SVAPRA12} J. Sperling, W. Vogel, and G. S. Agarwal, Phys. Rev. A {\bf 85}, 023820 (2012).
	\bibitem{SVAPRL12} J. Sperling, W. Vogel, and G. S. Agarwal, Phys. Rev. Lett. {\bf 109}, 093601 (2012).
	\bibitem{BDJDBW13} T. J. Bartley, G. Donati, Xian-Min Jin, A. Datta, M. Barbieri, and I. A. Walmsley, Phys. Rev. Lett. {\bf 110}, 173602 (2013).
	\bibitem{BDFL08} J.-L. Blanchet, F. Devaux, L. Furfaro, and E. Lantz, Phys. Rev. Lett. 101, 233604 (2008).
	\bibitem{MMDL12} P.-A. Moreau, J. Mougin-Sisini, F. Devaux, and E. Lantz, Phys. Rev. A {\bf 86}, 010101(R) (2012).
	\bibitem{SVAPRA13} J. Sperling, W. Vogel, and G. S. Agarwal, Phys. Rev. A {\bf 88}, 043821 (2013).
	\bibitem{PHMH12} J. Pe\v{r}ina, Jr., O. Haderka, V. Mich\'{a}lek, and M. Hamar, Opt. Lett. {\bf 37}, 2475 (2012).
	\bibitem{PHMH13} J. Pe\v{r}ina, Jr., O. Haderka, V. Mich\'{a}lek, and M. Hamar, Phys. Rev. A {\bf 87}, 022108 (2013).
	\bibitem{AT92} G. S. Agarwal and K. Tara, Phys. Rev. A {\bf 46}, 485 (1992).
	\bibitem{ZVB04} A. Zavatta, S. Viciani, and M. Bellini, Science {\bf 306}, 660 (2004).
	\bibitem{BA07} A. Biswas and G. S. Agarwal, Phys. Rev. A {\bf 75}, 032104 (2007).
	\bibitem{KVPZB08}  T. Kiesel, W. Vogel, V. Parigi, A. Zavatta, and M. Bellini, Phys. Rev. A {\bf 78}, 021804(R) (2008).
	\bibitem{DM09} A. V. Dodonov and S. S. Mizrahi, Phys. Rev. A {\bf 79}, 023821 (2009).
	\bibitem{KVBZ11} T. Kiesel, W. Vogel, M. Bellini, and A. Zavatta, Phys. Rev. A {\bf 83}, 032116 (2011).
	\bibitem{LLNJ12} C.-W. Lee, J. Lee, H. Nha, and H. Jeong, Phys. Rev. A {\bf 85}, 063815 (2012).
	\bibitem{FMCZB13} S. N. Filippov, V. I. Man'ko, A. S. Coelho, A Zavatta, and M Bellini, Phys. Scr. {\bf T153}, 014025 (2013).
	\bibitem{RKVGZB13} S. Rahimi-Keshari, T. Kiesel, W. Vogel, S. Grandi, A. Zavatta, and M. Bellini, Phys. Rev. Lett. {\bf 110}, 160401 (2013).
	\bibitem{MF10} P. Marek and R. Filip, Phys. Rev. A {\bf 81}, 022302 (2010).
	\bibitem{F09} J. Fiur\'{a}\v{s}ek, Phys. Rev. A {\bf 80}, 053822 (2009).
	\bibitem{PZKB07} V. Parigi, A. Zavatta, M. Kim, and M. Bellini, Science {\bf 317}, 1890 (2007).
	\bibitem{C82} C. M. Caves, Phys. Rev. D {\bf 26}, 1817 (1982).
	\bibitem{RL09} T. C. Ralph and A. P. Lund, in {\it Quantum Communication Measurement and Computing Proceedings of 9th International Conference}, edited by A. Lvovsky (AIP, New York, 2009). 
	\bibitem{FBBFTG10} F. Ferreyrol, M. Barbieri, R. Blandino, S. Fossier, R. Tualle-Brouri, and P. Grangier, Phys. Rev. Lett. {\bf 104}, 123603 (2010).
	\bibitem{ZFB11} A. Zavatta, J. Fiur\'{a}\v{s}ek, and M. Bellini, Nature Photon. {\bf 5}, 52 (2011).
	\bibitem{BPMT13} N. Bruno, V. Pini, A. Martin, and R. T. Thew, New J. Phys. {\bf 15}, 093002 (2013).
	\bibitem{MD02} S. S. Mizrahi and V. V. Dodonov, J. Phys. A: Math. Gen. {\bf 35}, 8847 (2002).
	\bibitem{BCZW10} S. Barz, G. Cronenberg, A. Zeilinger, and P. Walther, Nature Photon. {\bf 4}, 553 (2010).
	\bibitem{HHFRRJ10} H. H\"ubel, D. R. Hamel, A. Fedrizzi, S. Ramelow, K. J. Resch, and T. Jennewein, Nature {\bf 466}, 601 (2010).
	\bibitem{XRLWP10} G. Y. Xiang, T. C. Ralph, A. P. Lund, N. Walk, and G. J. Pryde, Nature Photon. {\bf 4}, 316 (2010).
	\bibitem{LN12} S.-Y. Lee and H. Nha, Phys. Rev. A {\bf 85}, 043816 (2012).
	\bibitem{NGSC12} C. Navarrete-Benlloch, R. Garc\'{i}a-Patr\'{o}n, J. H. Shapiro, and N. J. Cerf, Phys. Rev. A {\bf 86}, 012328 (2012).
	\bibitem{OBSZGT12} C. I. Osorio, N. Bruno, N. Sangouard, H. Zbinden, N. Gisin, and R. T. Thew, Phys. Rev. A {\bf 86}, 023815 (2012).
	\bibitem{AN13} U. L. Andersen and J. S. Neergaard-Nielsen, Phys. Rev. A {\bf 88}, 022337 (2013).
	\bibitem{GLS13} R. Ghobadi, A. Lvovsky, and C. Simon, Phys. Rev. Lett. {\bf 110}, 170406 (2013).
	\bibitem{MBHSAGLS13} O. Morin, J.-D. Bancal, M. Ho, P. Sekatski, V. D’Auria, N. Gisin, J. Laurat, and N. Sangouard, Phys. Rev. Lett. {\bf 110}, 130401 (2013).
	\bibitem{SHRSHB13} L. Slodi\u{c}ka, G. H\'{e}tet, N. R\"ock, P. Schindler, M. Hennrich, and R. Blatt, Phys. Rev. Lett. {\bf 110}, 083603 (2013).
	\bibitem{BookVogel} W. Vogel and D.-G. Welsch, \textit{Quantum Optics}, (Wiley-VCH, Weinheim, 2006).
	\bibitem{BookAgarwal} G. S. Agarwal, \textit{Quantum Optics}, (Cambridge University Press, Cambridge, 2013).
	\bibitem{ASV13} E. Agudelo, J. Sperling, and W. Vogel, Phys. Rev. A {\bf 87}, 033811 (2013).
	\bibitem{GSV12} C. Gehrke, J. Sperling, and W. Vogel, Phys. Rev. A {\bf 86}, 052118 (2012).
	\bibitem{G63} R. J. Glauber, Phys. Rev. Lett. {\bf 10}, 84 (1963).
	\bibitem{S63} E. C. G. Sudarshan, Phys. Rev. Lett. {\bf 10}, 277 (1963).
	\bibitem{ASW93} G. S. Agarwal, M. O. Scully, and H. Walther, Phys. Scr. {\bf T48}, 128 (1993).
	\bibitem{PJCC13} S. Pandey, Z. Jiang, J. Combes, and C. M. Caves, Phys. Rev. A {\bf 88}, 033852 (2013).
	\bibitem{AP07} G. S. Agarwal and P. K. Pathak, Phys. Rev. A {\bf 75}, 032108 (2007).
\end{thebibliography}
\end{document}